# Why Doesn't Microsoft Let Me Sleep? How Automaticity of Windows Updates Impacts User Autonomy

Sanju Ahuja

Indian Institute of Technology Delhi, sanju.ahuja@design.iitd.ac.in

Ridhi Jain

Indian Institute of Technology Delhi, ridhijain821@gmail.com

Jyoti Kumar

Indian Institute of Technology Delhi, jyoti@design.iitd.ac.in

'Automating the user away' has been designated as a dark pattern in literature for performing tasks without user consent or confirmation. However, limited studies have been reported on how users experience the sense of autonomy when digital systems fully or partially bypass consent. More research is required to understand what makes automaticity a threat to autonomy. To address this gap, a qualitative interview study with 10 users was conducted to investigate the user experience of Microsoft Windows updates. It was found that ten design features of Windows updates impact the autonomy experience. For each design feature, the contextual factors which influence its impact on autonomy were also noted. The findings of this paper can help designers understand the ethical concerns posed by automaticity in design and identify measures to mitigate these concerns.

**CCS CONCEPTS** • Human-centered computing ~ Human computer interaction (HCI) ~ Empirical studies in HCI

**Additional Keywords and Phrases:** dark patterns, automating the user, user experience, autonomy, consent

## 1 INTRODUCTION

In computing systems, the term 'dark pattern' is used to refer to a design feature which seeks to influence users into acting in certain ways, often against their intentions or their best interests [1, 2]. Gray et al. [3] identified 'automating the user away' as a dark pattern, referring to design features that automate the process of performing tasks without the consent or confirmation of the user. In such designs, automated actions are performed, often at critical moments, without any warning or an opportunity for the user to consent. The automaticity of computing systems has emerged as an ethical concern in different contexts such as privacy, proxemic interactions and e-commerce, where users are subject to designs which either bypass consent altogether or assume consent by default [4-6].

However, not all automatic system behaviors can be designated as unethical. Automaticity often serves functional purposes and users may not always desire complete control over every aspect of the system. It remains under explored what makes automaticity problematic from a user autonomy perspective. User studies of dark patterns have primarily focused on 'manipulative' designs [7-9], and relatively less on design which 'automates the user away'. To address this gap, this paper aims to understand how automaticity impacts the experience of autonomy. In Gray et al.'s findings [3], Microsoft Windows updates emerged as a prominent example of automaticity in design which was criticized by users. Hence, this paper selected Windows updates as a use case to investigate the impact of automatic system behaviors on autonomy. For this purpose, a qualitative interview study was conducted with 10 users to understand the experience of interacting with a Windows update.

## 2 RESEARCH METHODOLOGY

### 2.1 Participants and Study Design

Ten participants were invited for the study on a voluntary basis. The mean age of the participants was 22.9 years (SD = 1.2 years). Six participants identified as female and four identified as male. Two participants were undergraduate students, one was a postgraduate student, seven had university degrees out of whom six were employed. Participants' consent to take part in the study and to record the interviews was collected via a digital consent form. The average duration of an interview was 17.1 minutes.

### 2.2 Interview Protocol

The interview protocol was created to understand participants' experiences of interacting with Windows updates without explicitly probing them about dark patterns or nudging them to think about these updates as unethical. This was done to reduce bias in participants' responses. Within the interview protocol, recall-based questions such as *'Do you remember when an update last happened on your computer?'* were only included to guide the conversation, and not for frequency analysis. The interview protocol was as follows:

a) Do you use a Microsoft Windows computer?
b) Have you ever had your Windows update automatically?
c) How frequently does a Windows update happen?
d) Do you remember when an update last happened on your computer?
e) Does the update happen automatically or is there a warning or a prompt?
f) Did you ever lose any important data? How did you feel about that? Do you take any steps to prevent data loss?
g) Have you missed any important meetings or deadlines because of the update? At that time, do you blame Microsoft or the designer, or do you personally feel responsible?
h) Are there any other devices or software in which you face automatic updates?

### 2.3 Data Analysis

The interviews were transcribed for analysis. At first, the transcripts were open coded in NVivo to capture various elements of the participants' responses. Next, equivalent codes were combined and similar codes were clustered. To identify design features of the Windows update which impacted users' autonomy, interview segments were examined where participants narrated their autonomy related experiences. These segments included both a direct verbalization of the experience of autonomy (Ex. 'we don't have a choice') as well as indirect references (Ex. 'why does this Microsoft is like not letting me sleep?'). From these segments, design features which negatively impacted autonomy were identified. It was also observed that the autonomy impact of each design feature was influenced by contextual factors. These factors could either prevent the feature from being experienced as an autonomy threat (positive influence) or they could contribute to it (negative influence). The data was then subjected to a thematic analysis [10]. The formation of themes followed a top down approach, based on how the meaning of autonomy has been understood in literature [11, 12]. This paper reports three themes in the data and each theme contains features which impact different aspects of user autonomy.

## 3 RESULTS

All participants in the study (n=10) were current users of a Microsoft Windows computer with prior experience with Windows updates, albeit with different frequencies. The findings suggest that different features of Windows updates can undermine different aspects of autonomy. The three kinds of autonomy experiences that emerged from the data pertain to *freedom of choice, control* and *agency*. Figure 1 reports a summary of design features which impact the autonomy experience in each sense, and the number of participants who reported these design features. It also reports the contextual factors which influence the autonomy impact of each design feature.

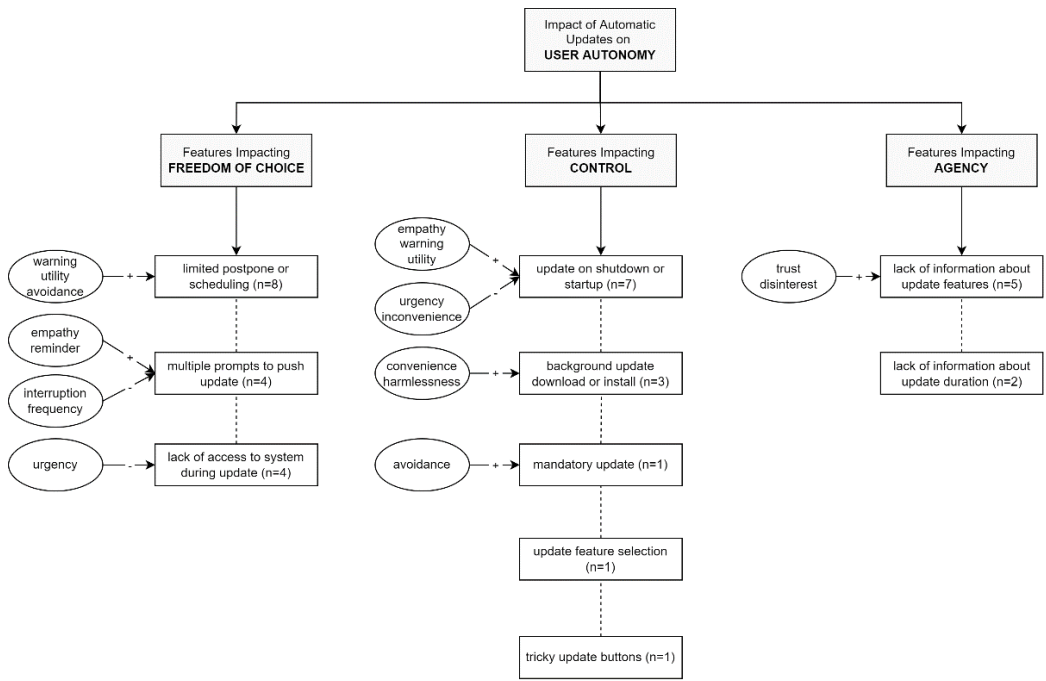

Figure 1: Design features and contextual factors which impact the autonomy experience of Windows updates

### 3.1 Features Impacting 'Freedom of Choice'

According to literature, freedom of choice is concerned with choices or options available to an individual [11, 12]. Freedom can be threatened by limiting users' choices or by pressuring them into choosing certain options [13]. With this understanding, it was found that three design features of automatic updates negatively impact users' freedom. These are:
  a) limited choices to postpone or reschedule the update (such as how many times the update can be postponed or for how many days it can be rescheduled),
  b) multiple prompts to push the update to the user, and
  c) an inability to access the system during the update or to quit the update

Figure 2 reports participants' responses about limited choices to postpone or schedule the update (n=8). They report experiences such as 'not having a choice', 'not having options' and 'not having any say', indicating a negative impact on freedom of choice. However, there were contextual factors which influenced the impact of each feature on the participants' experience. For example, in Figure 2, P2's responses indicate that if there is sufficient warning, then limited choices to postpone the update is an acceptable restriction and it does not feel 'forced'. P3's responses indicate a tolerance towards this feature because he feels that it is necessary to update his system.

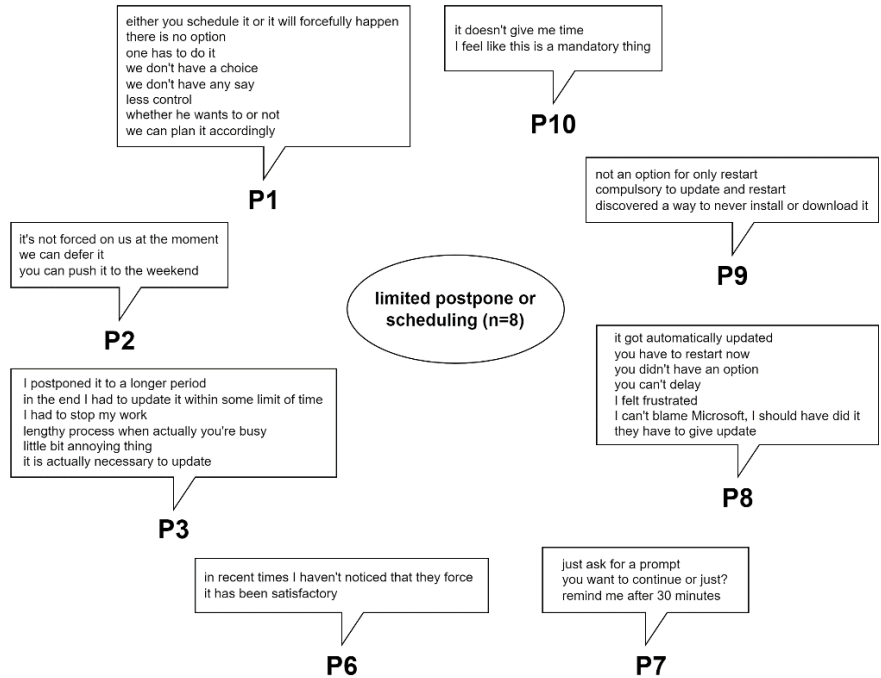

Figure 2: Autonomy experience reports pertaining to a single design feature ('limited postpone or scheduling', n=8)

### 3.2 Features Impacting 'Control'

Control can be understood as a user having the opportunity to act in accordance to their preferences and to consent to the actions of a system as opposed to the system acting on their behalf [11]. When control is lost, it means that decisions are made for users without their active participation. From the interviews, five design features of Windows updates were identified as impacting users' experience of control. These are:

a) the update automatically happening on system startup or shutdown,
b) the update being installed in the background without users' knowledge,
c) the update being mandatory such that the user cannot avoid installing it eventually,
d) lack of opportunities to select specific features of the update, and
e) tricky update buttons which the user can click by mistake

These features were themed together because they redirect the decision making fully or partially to the system, without involving the user. This suggests that the system is deemed more capable, adept, or intelligent at choosing for the user [3], undermining their desire and their experience of control, as indicated by one of the responses below.

- P5: *"I think it should not start automatically like when I restart my computer so suddenly it should not like start installing updates because if I have to do some work, then I have to wait until everything my computer restarts. I think it should ask the user what will be a good time to restart every time."*

However, there are contextual factors which can influence (positively or negatively) how each feature impacts users' experience of control (Figure 1). For example, if users are empathetic towards the designer, or they believe that they have been given sufficient warning, or that the updates are useful, then the experience of control is not strongly undermined (P2). These are positive influences. On the contrary, when users need urgent access to their system, their experience of control is negatively impacted. This is a negative influence on control.

### 3.3 Features Impacting 'Agency'

In literature, agency has been understood as the ability of an individual to evaluate or reason about a decision in accordance with their goals, preferences and values [11, 12]. Agency can be undermined through the lack of relevant information, or through intentionally manipulative tricks and strategies [2]. In the interviews, two design features of Windows updates were identified as impacting agency. These are:

a) lack of adequate information about the features of the update, and
b) lack of information about how long the update will take

These deficiencies prevent users from adequately planning the update around their schedule. For some users, they also prevent users from deciding whether they actually want the update in their system or not. For example, the following quote from P9 shows his inability to decide whether the update is beneficial for his system, because relevant information is not provided before the update.

- P9: *"There should be maybe a rating or something like, the user experience of that, like what user are facing issues, what user users like. Is it stable or not, like what are the problems they are facing? It is worth installing or not? Like something like that."*

However, two contextual factors seem to alleviate these concerns: disinterest and trust (Figure 1). For example, P8 mentions that he is not interested in receiving information about the update, and hence lack of this information does not appear to impact his sense of agency. However, he expresses a desire to receive information about the update duration. P4 says that he trusts Microsoft to do its job, so he does not care so much about having more information about the update.

## 4 DISCUSSION AND FUTURE WORK

This study took the approach of a user experience investigation to identify design features of Microsoft Windows updates which impact user autonomy. While user perceptions of dark patterns have been studied previously in literature [7-9], 'automaticity' as a dark pattern has not received sufficient attention. This study addresses this gap. The findings revealed the role of several individual and contextual factors in assessing whether or not automating the user away is a threat to autonomy. User interviews showed that the mere presence of automaticity is not sufficient for a design to be experienced as problematic. The findings of this study can help designers understand the distinct nature of the autonomy concerns posed by the various features of automatic updates. They can also point to mitigative solutions, such as providing sufficient warning, clearly communicating the usefulness and the duration of updates, providing an opportunity to bypass the updates whenever possible, providing an opportunity to quit the updates if they take too long, and not disrupting users during ongoing tasks. These suggestions may of course be somewhat subject to technical viability. Nevertheless, the findings do show the importance of understanding users' needs and context while designing features that fully or partially bypass their consent.